\documentclass[aps,twocolumn,superscriptaddress,showpacs,floats]{revtex4}
\usepackage{amsmath}
\usepackage{amsfonts}
\usepackage{amssymb}
\usepackage{euscript}
\usepackage{bm}
\begin{document}

\title{Comment on ``Dispersive bottleneck delaying thermalization of turbulent Bose-Einstein Condensates'' by Krstulovic and Brachet}

\author{Evgeny Kozik}
\affiliation{Institute for Theoretical Physics, ETH Zurich, CH-8093 Zurich, Switzerland}

%


\maketitle

The Letter by Krstulovic and Brachet (KB) \cite{numerics} addresses an important problem in kinetics of Bose-Einstein
condensation (BEC) in a weakly interacting Bose gas (WIBG). By means of a 
numeric simulation of the truncated Gross-Pitaevskii equation (TGPE) (the high-wave-number harmonics are cut off at $k_\mathrm{max}$), the authors observe a peculiar relaxation picture: starting with a superfluid vortex state, late-time evolution of the energy distribution towards thermal equilibrium takes on the form of a front at a characteristic wave number $k_c(t)$ (at which the energy is concentrated) with an abrupt truncation at the wave numbers $k_\mathrm{max}> k>k_c$ \cite{TGPE}. The front propagates toward higher $k$'s at an ever-decreasing rate $\dot{k}_c(t)$ leaving a quasithermalized distribution in its wake at $k<k_c$. The Letter puts forward ``a new mechanism of thermalization'' suggesting that ``a bottleneck delays the final thermalization when large dispersive effects
are present at truncation wave number and produces an effective self-truncation.'' We point out that the physical mechanism responsible for the numeric results is in fact quite different \cite{Sv91}: (i) in contrast to the bottleneck proposed in \cite{numerics}, the delay in thermalization is due to the smooth increase of the typical kinetic time $\tau_\mathrm{kin}(k)$ with the wave number $k$, namely $\tau_\mathrm{kin}(k) \propto k$, correspondingly (ii) the effective self-truncation is a consequence of energy conservation. The observed physics is well understood as a part of the general relaxation scenario of a strongly nonequilibrium WIBG developed by Svistunov \cite{Sv91}, describing the underlying kinetic mechanisms, the form of the distribution, and its time dependence as $k_c(t) \propto t^{1/4}$, which the numerics [Fig.~4(e) of Ref.~\cite{numerics}] agree with.

The general relaxation scenario \cite{Sv91} is nontrivial starting with an explosive wave towards $k=0$, which populates long-wavelength harmonics and leads to the formation of the quasicondensate (QC)---the superfluid turbulence state, i.e. a tangle of quantized vortices---and a subsequent back wave propagating towards large $k$'s with a quasiequilibrium distribution formed behind the front. The simulation \cite{numerics} corresponds to the late-time stage of this scenario, in which most of the particles are already in the QC (represented in \cite{numerics} by a decaying Taylor-Green vortex). This regime is described by the kinetic equation (KE) for the mode occupation numbers $n_k$, 
\begin{equation}
\dot{n_k}\;=\;N_0\, \mathrm{Coll}_0(\,[n_k], k\,), \label{KE}
\end{equation}
where $\mathrm{Coll}_0 \propto k  \,n_k^2 $ is the collision integral and $N_0$ is the number of QC particles, the specifics of QC dynamics being irrelevant \cite{footnote}. $N_0$ is close to the total particle number and can be considered constant. Physically, the KE describes the dominant three-wave collision processes with the fourth wave being the QC $N_0$. Scale invariance of the KE along with the conservation of the total energy $E \propto \int k^4 dk \, n_k$ dictates that the solution describing the thermalization at high $k$'s takes on a self-similar form (Eq.~(4.4) in Ref.~\cite{Sv91}):
\begin{equation}
n_k\;=\;k_c(t)^{-5} f(k/k_c(t)). \label{solution}
\end{equation}
Here $f(x) \to 0, \; x \gg 1$ enforcing truncation at $k_c$ to ensure convergence of $E$, and $f(x) \propto 1/x^2, \; x \ll 1$ corresponds to the Gibbs distribution for $k \ll k_c$. The evolution of $k_c(t)$ follows from the KE~(\ref{KE}) yielding $k_c(t) \propto t^{1/4}$. 

Eqs.~(\ref{KE}),~(\ref{solution}) prescribe that kinetics get gradually slower with $k$, $\tau_\mathrm{kin}(k) \propto k$.  That is the fundamental reason for the deceleration of the front propagation at $k_c(t)$ and the thermal equilibration of the modes in its wake. The solution~(\ref{solution}) describes a drift of energy [contrasted with an energy cascade, in which $\tau_\mathrm{kin}(k)$ vanishes towards $k \to \infty$) \cite{Sv91}, whereas a bottleneck, i.e. an abrupt drop in kinetic efficiency at some $k$-scale (typically due to a change of the kinetic mechanism], does not take place.

KB also raise an interesting question of whether the ultimate thermalization is completely inhibited in the limit of an arbitrarily high spatial resolution due to the ever-slowing-down nature of the process at high $k$'s. In real systems, the classical-field regime of the GPE breaks down at the wave numbers for which $n_k \sim 1$ and the kinetics are dominated by the quantum spontaneous (as opposed to stimulated at $n_k \gg 1$) scattering processes \cite{Sv91}. The spontaneous scattering provides an efficient mechanism for the equilibration at high wave numbers.

\end{document}